\documentclass{article}

\usepackage{arxiv}

\usepackage{amssymb}
\usepackage{amsmath}

\usepackage{tikz}
\usetikzlibrary{arrows.meta,positioning,fit,calc,shapes.misc}
\usepackage{pgfplots}
\usepackage{pgfplotstable}
\pgfplotsset{compat=1.18}
\usepackage{booktabs}
\usepackage{tabularx}
\usepackage{multirow}
\usepackage{array}
\usepackage{url}
\usepackage{hyperref}
\usepackage{algorithm}
\usepackage{orcidlink}
\usepackage{algpseudocode}
\usepackage{tikz}
\usetikzlibrary{arrows.meta,positioning,calc,fit,backgrounds,shadows.blur,shapes.geometric}
\usepackage{graphicx}

\newcommand{\AuthorBio}[3]{%
\vspace{1em}
\noindent
\begin{minipage}[t]{0.16\textwidth}
    \includegraphics[width=\linewidth]{#1}
\end{minipage}\hfill
\begin{minipage}[t]{0.82\textwidth}
    \textbf{#2} #3
\end{minipage}
\par
}

\definecolor{bwBox}{HTML}{FFFFFF}
\definecolor{bwStroke}{HTML}{000000}
\definecolor{bwMuted}{HTML}{4B5563}
\definecolor{bwBg}{HTML}{F9FAFB}

\title{VineetVC: Adaptive Video Conferencing Under Severe Bandwidth Constraints Using Audio-Driven Talking-Head Reconstruction}

\author{
Vineet Kumar Rakesh~\orcidlink{0009-0000-7102-6564} \\
Engineering Sciences, Homi Bhabha National Institute \\
Training School Complex, Anushaktinagar, Mumbai, Maharashtra 400094, India \\
Computer and Informatics Group, Variable Energy Cyclotron Centre \\
1/AF, Bidhannagar, Kolkata, West Bengal 700064, India \\
\texttt{vineet@vecc.gov.in} \\
\And
Soumya Mazumdar~\orcidlink{0009-0006-3521-9557} \\
Department of Computer Science and Business Systems \\
Gargi Memorial Institute of Technology \\
Baruipur, Kolkata, West Bengal 700144, India \\
\texttt{reachme@soumyamazumdar.com} \\
\And
Tapas Samanta~\orcidlink{0000-0003-0521-0747} \\
Computer and Informatics Group, Variable Energy Cyclotron Centre \\
1/AF, Bidhannagar, Kolkata, West Bengal 700064, India \\
Engineering Sciences, Homi Bhabha National Institute \\
Training School Complex, Anushaktinagar, Mumbai, Maharashtra 400094, India \\
\texttt{tsamanta@vecc.gov.in} \\
\And
Hemendra Kumar Pandey~\orcidlink{0000-0001-7203-2990} \\
Radioactive Ion Beam Facilities Group, Variable Energy Cyclotron Centre \\
1/AF, Bidhannagar, Kolkata, West Bengal 700064, India \\
Engineering Sciences, Homi Bhabha National Institute \\
Training School Complex, Anushaktinagar, Mumbai, Maharashtra 400094, India \\
\texttt{hkpandey@vecc.gov.in} \\
\And
Amitabha Das~\orcidlink{0009-0003-1460-8308} \\
School of Nuclear Studies and Application \\
Jadavpur University \\
Salt Lake City, Kolkata, West Bengal 700106, India \\
\texttt{amitabhad.snsa@jadavpuruniversity.in} \\
\And
Sarbajit Pal~\orcidlink{0009-0009-5246-7052} \\
Engineering Sciences, Homi Bhabha National Institute \\
Training School Complex, Anushaktinagar, Mumbai, Maharashtra 400094, India \\
\texttt{sarbajit@vecc.gov.in} \\
}

\begin{document}
\maketitle
\newpage
\begin{abstract}
Intense bandwidth depletion within consumer and constrained networks has the potential to undermine the stability of real-time video conferencing: encoder rate management becomes saturated, packet loss escalates, frame rates deteriorate, and end-to-end latency significantly increases. This work delineates an adaptive conferencing system that integrates WebRTC media delivery with a supplementary audio-driven talking-head reconstruction pathway and telemetry-driven mode regulation. The system consists of a WebSocket signaling service, an optional SFU for multi-party transmission, a browser client capable of real-time WebRTC statistics extraction and CSV telemetry export, and an AI REST service that processes a reference face image and recorded audio to produce a synthesized MP4; the browser can substitute its outbound camera track with the synthesized stream with a median bandwidth of 32.80 kbps. The solution incorporates a bandwidth-mode switching strategy and a client-side mode-state logger.
\end{abstract}


\section{Introduction}

Video calls in real time are now a significant tool for business, school, and health care. During the COVID-19 pandemic, telework, teleconferencing, and online learning became much more common, and many of these habits stayed the same after the pandemic \cite{mouratidis2021telework}. Telemedicine grew as well, and video consultations became a regular means to contact patients when it was hard to see them in person \cite{omboni2022telemedicine}. Because of this, more and more consumers want video conversations to perform well over a wide range of networks, such as mobile and shared home connections.

Video conferencing, on the other hand, still stops working when the bandwidth is extremely low or inconsistent. Real-world tests of popular platforms reveal that "comfortable" quality generally requires multi-megabit uplink/downlink speeds. Latency and packet loss may rapidly make video blurry, freeze, and make audio and video not sync up well \cite{netforecast2021vcrequirements}. Network problems are particularly bad for compressed video since missing even a few packets might make it impossible to decode not just the current frame but also future frames (because of inter-frame prediction), which produces freezing that can be seen \cite{rudow2023tambur}. These challenges happen more often in rural or distant places and in developing countries where bandwidth is limited and mobile phone coverage might change.

WebRTC is the basis for most browser-based video conferencing solutions. It lets media travel between peers, get across NAT, and be delivered securely in real time \cite{webrtcW3C}. WebRTC usually employs good audio codecs (like Opus) and video codecs (like VP8/H.264/VP9/AV1, depending on how it is set up). Opus is designed for interactive communication and can work with extremely low bitrates while still making speech understandable \cite{rfc6716}. WebRTC also has a standard statistics interface called getStats that lets apps keep track of things like bitrate, loss, latency, and other important call indicators \cite{webrtcStatsW3C}. Selective Forwarding Units (SFUs) are a standard way to increase distribution for multi-party calls without having to re-encode everything \cite{rfc7667}. Even with these improvements, traditional systems still need to communicate compressed video frames. When the available bandwidth goes below a few hundred kbps, the quality and continuity of the video frequently become worse quickly.

A hopeful alternative is to think of "video presence" as something that can be built up instead than sent. Recent audio-driven talking-head techniques may produce authentic mouth movement (and sometimes head movement) using a reference face and a speech signal \cite{prajwal2020wav2lip,zhou2020makeittalk,zhang2023sadtalker}. This provides a bandwidth-saving strategy: instead of continually delivering pixel video, the system may broadcast low-bitrate audio with a concise description of face movements and rebuild a talking-head stream at the receiver.  Since the payload is substantially less than full video, such a mode may remain useable even under severe bandwidth restrictions. 

In this research, we introduce (Versatile Intelligent Network Enhanced Efficient Telepresence), an adaptive video conferencing platform that blends WebRTC transmission with an auxiliary audio-driven reconstruction route.  The core notion is simple: while the network is robust, the call operates as a conventional WebRTC video session; when bandwidth collapses, the system changes to an AI reconstruction mode to retain a video-like experience while keeping the sent rate very low.  The mode selection is determined by real-time telemetry acquired via the WebRTC statistics API \cite{webrtcStatsW3C}.  In our prototype, the client may substitute the outgoing camera track with a synthetic talking-head stream derived from the user’s reference face and voice.  This approach targets the severe-bandwidth regime, when additional video compression delivers decreasing returns and user experience becomes dominated by freezes and instability \cite{rudow2023tambur}. Concretely, our current deployment integrates several state-of-the-art audio-driven talking-head generators behind a uniform service API. Unless otherwise stated, all quantitative results in this paper use Wav2Lip\cite{prajwal2020wav2lip} as the default backend, which is a strong and widely adopted lip-sync model. To demonstrate that our system is not tied to a single architecture, we have also implemented backends for MakeItTalk\cite{zhou2020makeittalk}, SadTalker\cite{zhang2023sadtalker}, and VadicTHG model, which can be switched at runtime without changing the rest of the video conferencing pipeline. This work connects two research directions: (i) how mainstream video conferencing systems behave under constrained network conditions, and (ii) how talking-head generation can provide a video-like experience without continuously transmitting full video frames.

The main contributions of this work are:
\begin{itemize}
    \item \textbf{Adaptive conferencing design:} a WebRTC-based platform with telemetry-driven mode control using standardized browser statistics \cite{webrtcW3C,webrtcStatsW3C}.
    \item \textbf{Ultra-low bitrate ``video presence'' mode:} an audio-driven talking-head reconstruction path, motivated by recent advances in lip-sync and talking-head generation \cite{prajwal2020wav2lip,zhou2020makeittalk,zhang2023sadtalker}.
    \item \textbf{Engineering implementation:} a working prototype with signaling, optional SFU support (RTP topologies), and client-side track replacement for synthesized video \cite{rfc7667}.
    \item \textbf{Evidence under real constraints:} long-run bandwidth logging that quantifies the bandwidth gap between normal video and AI reconstruction mode (reported in the Results section and summarized in Table~\ref{tab:bandwidth_summary}).
\end{itemize}

\section{Related Work}
\label{sec:related_work}

A primary difficulty in bandwidth-constrained video conferencing is the lack of uniformity across platforms, which vary in media stacks, codec setups, congestion management mechanisms, server-side rules, and the degree of transparency provided to users and researchers. The variability complicates cross-platform findings unless studies are structured for reproducibility, using a common stimulus and a standardized set of measurements \cite{ZHANG2025112418}. A new bandwidth-aware benchmarking research demonstrates that many systems stay stable under moderate-to-good network settings, but quality becomes extremely variable as bandwidth declines and adaption strategies diverge. In very limited environments, platforms may be unable to initiate sessions, deactivate video, or revert to audio-only functionality, revealing realistic lower thresholds that are often obscured from user-facing specifications \cite{parsonson-2022}. Importantly, the study highlights that behavior at high available bandwidth can still be limited by vendor-imposed policy caps rather than network constraints, so interpreting results requires separating policy-driven behavior from true network-driven adaptation \cite{megyesi-2016}. The study highlights that time-resolved indicators, such as frame rate versus time and resolution versus time, provide greater insight into user-perceived degradation than throughput alone. Additionally, extreme scenarios may necessitate categorical outcomes alongside continuous metrics \cite{turner-2010}. Modern browser-based conferencing solutions generally depend on WebRTC, which offers a standardized statistics interface that exposes transport- and RTP-level metrics via browser APIs. The benchmarking study highlights that the WebRTC statistics model (via \texttt{getStats}) provides a viable framework for collecting similar measures across browser-based apps, including throughput trends, frame dynamics, jitter, and packet-loss indicators. It also highlights that observability vary by client modality: browser implementations often give richer, standardized metrics than native clients, which may show only partial diagnostics or obscure internal adaption state \cite{riveiro-2021}. These restrictions inspire system architectures that (i) can employ telemetry that is dependably accessible at the client, and (ii) can ensure conversational continuity even when standard rate adaptation is no longer adequate. Talking-head generation (THG) has evolved as a feasible approach to portray ``visual presence'' without continually broadcasting camera footage. The THG survey organizes prior work by input modality and generation paradigm, covering approaches that animate a face from images, audio, text, or driving videos, as well as methods built on keypoint motion transfer, explicit three-dimensional modeling, neural rendering, diffusion-based generation, and animation pipelines \cite{rakesh_thg_survey}. Across these families, a constant objective is to retain individuality while providing temporally coherent motion that coincides with speech. A clear trend highlighted in the survey is the movement from purely two-dimensional motion transfer toward three-dimensional and neural-rendering-based representations that better handle pose changes and improve realism, and further toward hybrid designs that combine neural rendering with stronger generative priors for better temporal smoothness and controllability \cite{rakesh_thg_survey}. The study also stresses real-time rendering focused THG, where architectures are intended for interactive use cases and attempt to minimize latency while ensuring identity and synchronization \cite{choi-2026}. Evaluation in talking-head synthesis often concentrates on a short number of key criteria: identity preservation, visual quality, audio-visual synchronization, and the naturalness of motion. The survey underlines that both objective measurements and subjective assessments are employed, and that newly suggested perceptual measures attempt to better represent human perception than pixel-level similarity alone \cite{rakesh_thg_survey}. Synchronization-focused assessment is vital because even physically convincing faces might seem unnatural if lip movements deviate from speech \cite{rakesh_thg_survey}. For deployment, the survey discusses several recurring limitations: reliance on large pre-trained generators (with possible dataset-induced biases), difficulty with extreme head pose and occlusions, challenges in multilingual settings due to language- and dataset-skewed training data, and temporal drift or flicker in longer sequences \cite{rakesh_thg_survey}. It also underlines the necessity for responsible deployment techniques, including protections and explicit governance for synthetic media, particularly in interactive environments such as education and video communication \cite{mazumdar-2025}. Taken together, the VC benchmarking literature explains that typical conferencing stacks may degenerate into unstable behavior or hard fallbacks under severe restrictions, and that telemetry-informed, repeatable assessment is important to understand these regimes \cite{naser-2025}. In parallel, the THG literature reveals that audio-driven face reconstruction is increasingly capable of providing synchronized, identity-preserving facial video suited for interactive applications, but confronts practical hurdles in robustness and long-duration stability \cite{zhang-2024}. This paper builds on these insights by treating ``visual presence'' as an adaptive modality: when network conditions make pixel streaming unreliable, the system can switch semantics from transmitting video frames to transmitting speech (and lightweight control cues) and reconstructing a talking head, while using client-visible telemetry to guide stable operation.

\section{System Overview}
\label{sec:system_overview}

\begin{figure}[t]
\centering
\includegraphics[width=\linewidth]{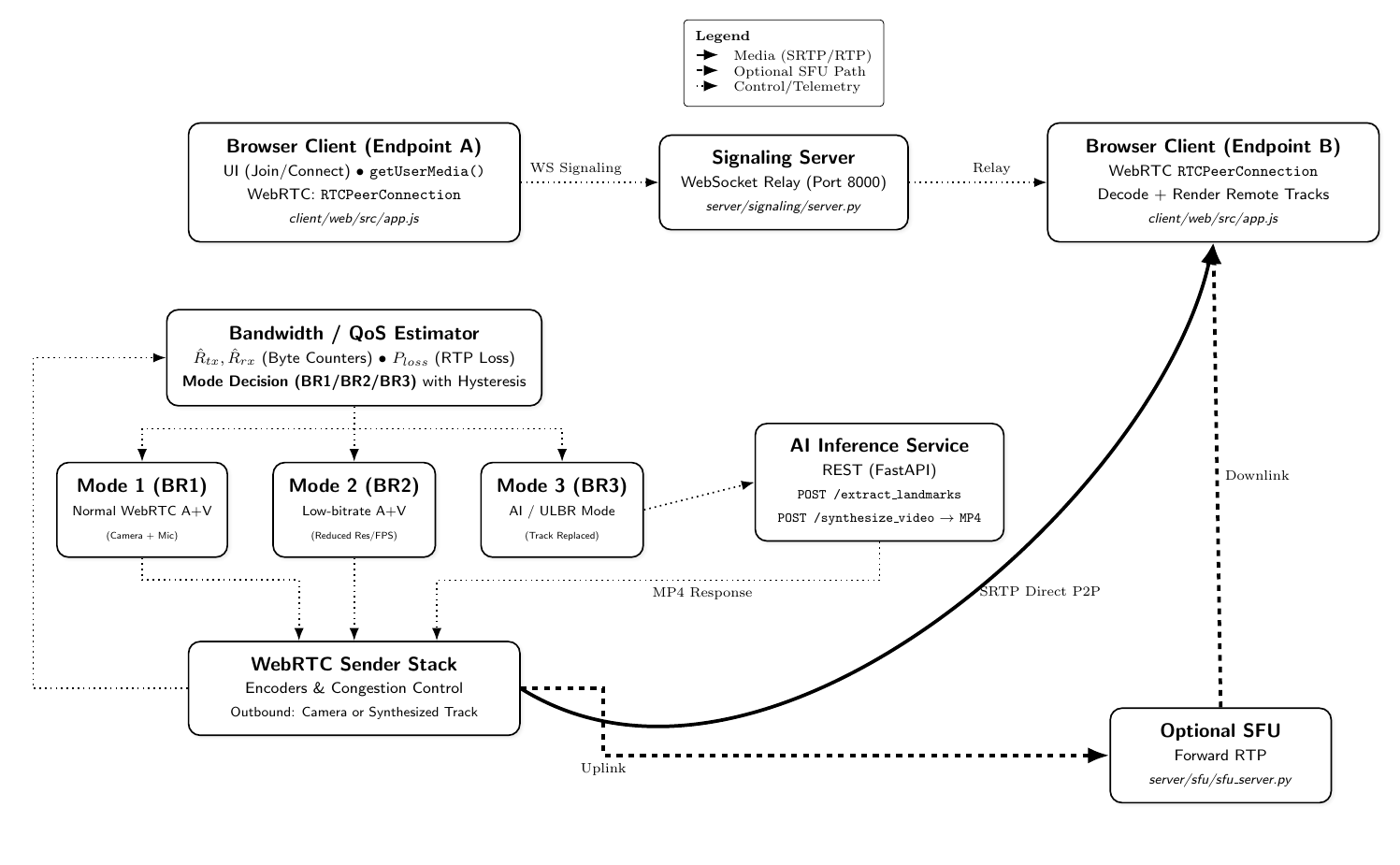}
\caption{System overview of the proposed bandwidth-adaptive video conferencing framework. After browser-based signaling and WebRTC session establishment, the sender periodically evaluates bandwidth/QoS and applies a three-mode policy: BR1 (standard A/V), BR2 (rate-constrained A/V), and BR3 (audio-only uplink with server-side talking-head synthesis and track replacement). Delivery is via SRTP P2P or an optional SFU.}
\label{fig:pipeline}
\end{figure}

A hybrid video-conferencing (VC) architecture is depicted in figure~\ref{fig:pipeline} that can operate in two complementary delivery paths: (i) a conventional WebRTC pixel-streaming path for normal network conditions, and (ii) an auxiliary audio-driven reconstruction path that replaces pixel video transmission with a reconstructible representation when bandwidth becomes severely constrained. The main purpose is to retain conversational continuity by modifying the semantics of what is delivered. Under acceptable bandwidth, the system functions like a normal VC application: camera and microphone are collected, encoded, and delivered in real time. Under severe bandwidth constraints, the system keeps the call active by continuing to transmit real-time audio while producing a synthetic talking-head video stream based on the audio and a face reference, so that a video-like presence is maintained even when sending continuous pixel video is no longer feasible. The media path relies on WebRTC for real-time transport. A signaling service exchanges the session descriptions and connectivity candidates required to establish a WebRTC session. This signaling channel carries only control information required for session setup and connectivity; after setup, audio and video media are carried as real-time RTP streams. For multi-party sessions, an SFU can be inserted to forward streams to multiple participants. In this case each endpoint uploads a single stream to the SFU, and the SFU selectively forwards the appropriate streams to other participants. This enables scalable group calls without requiring each sender to upload separate copies of its stream per receiver. A central element of the system is continuous telemetry extraction at the endpoint. Let $\Delta t$ denote the sampling interval of telemetry. The endpoint observes cumulative counters for the total bytes transmitted and received. Let $B_{\mathrm{tx}}(t)$ be the cumulative number of transmitted bytes up to time $t$ and $B_{\mathrm{rx}}(t)$ the cumulative number of received bytes up to time $t$. The instantaneous uplink and downlink throughput estimates in kbps are computed by differencing these counters over the most recent interval and converting bytes to bits:
\begin{equation}
\label{eq:tx_rate}
\hat{R}_{\mathrm{tx}}(t)=\frac{8\left(B_{\mathrm{tx}}(t)-B_{\mathrm{tx}}(t-\Delta t)\right)}{1000\,\Delta t},
\end{equation}
\begin{equation}
\label{eq:rx_rate}
\hat{R}_{\mathrm{rx}}(t)=\frac{8\left(B_{\mathrm{rx}}(t)-B_{\mathrm{rx}}(t-\Delta t)\right)}{1000\,\Delta t}.
\end{equation}
In \eqref{eq:tx_rate}--\eqref{eq:rx_rate}, $B_{\mathrm{tx}}(t)-B_{\mathrm{tx}}(t-\Delta t)$ is the number of bytes sent during the last interval of length $\Delta t$; multiplying by $8$ converts bytes to bits; dividing by $\Delta t$ converts the value to bits/s; finally dividing by $1000$ converts to kbps. These two estimates are sufficient to characterize how much network capacity is being used on each side of the call at time $t$ and provide a real-time signal to decide which operating mode is feasible. Throughput alone does not fully describe call stability under lossy links, so the endpoint also estimates loss-related indicators from RTP counters. Let $P_{\mathrm{lost}}(t)$ and $P_{\mathrm{recv}}(t)$ be cumulative counts of lost packets and received packets, respectively. The interval loss ratio is estimated as
\begin{equation}
\label{eq:loss_ratio}
\hat{p}_{\mathrm{loss}}(t)=
\frac{\Delta P_{\mathrm{lost}}(t)}{\Delta P_{\mathrm{lost}}(t)+\Delta P_{\mathrm{recv}}(t)},
\quad\text{where}\quad
\Delta P(t)=P(t)-P(t-\Delta t).
\end{equation}
This estimate is simple and easy to compute: it measures what fraction of packets in the last interval were lost. A useful capacity proxy that accounts for loss is the effective goodput, defined as the fraction of the transmitted bitrate that is expected to arrive successfully as per equation~\ref{eq:goodput}. The goodput $\hat{G}(t)$ decreases when packet loss increases, even if the sender attempts to transmit at the same rate. Since network measurements fluctuate, the controller uses a smoothed estimate to avoid unstable switching. A standard exponential moving average is applied based on the equation~\ref{eq:ewma}
\begin{equation}
\label{eq:goodput}
\hat{G}(t)=\hat{R}_{\mathrm{tx}}(t)\left(1-\hat{p}_{\mathrm{loss}}(t)\right).
\end{equation}
\begin{equation}
\label{eq:ewma}
\tilde{G}(t)=\alpha\,\tilde{G}(t-\Delta t)+(1-\alpha)\,\hat{G}(t),
\qquad 0<\alpha<1.
\end{equation}
Where $\alpha$ controls how quickly the estimate reacts to changes. If $\alpha$ is close to $1$, the estimate changes slowly and is robust to short spikes. If $\alpha$ is smaller, the estimate responds quickly to sudden drops. The system uses three operating modes, shown in equation~\ref{eq:ctr_mode}. Each mode is defined by what data is transmitted and therefore by a different decomposition of the total bitrate. Let the total transmitted bitrate be denoted by $R_{\mathrm{tot}}(t)$ and decomposed into four components which are mentioned in equation~\ref{eq:total_rate}.
\begin{equation}
\label{eq:total_rate}
R_{\mathrm{tot}}(t)=R_{\mathrm{aud}}(t)+R_{\mathrm{vid}}(t)+R_{\mathrm{ctrl}}(t)+R_{\mathrm{ref}}(t).
\end{equation}
In \eqref{eq:total_rate}, $R_{\mathrm{aud}}(t)$ is the audio bitrate, $R_{\mathrm{vid}}(t)$ is the bitrate of conventional compressed pixel video, $R_{\mathrm{ctrl}}(t)$ is the bitrate of compact control information used for reconstruction (for example, facial motion descriptors), and $R_{\mathrm{ref}}(t)$ is the average bitrate due to occasional reference updates. This decomposition is important because the proposed method improves performance under severe bandwidth constraints primarily by reducing $R_{\mathrm{vid}}(t)$ to near zero in AI mode while keeping $R_{\mathrm{aud}}(t)$ and low-overhead control terms. In the Normal mode, the system transmits conventional WebRTC audio and video, and does not rely on reconstruction signals:
\begin{equation}
\label{eq:normal_mode}
R_{\mathrm{tot}}^{(\mathrm{Normal})}(t)\approx R_{\mathrm{aud}}(t)+R_{\mathrm{vid}}(t),
\qquad
R_{\mathrm{ctrl}}(t)\approx 0,\;
R_{\mathrm{ref}}(t)\approx 0.
\end{equation}
This mode provides the highest visual quality because $R_{\mathrm{vid}}(t)$ is allocated to coded pixel video. However, as bandwidth decreases, maintaining $R_{\mathrm{vid}}(t)$ becomes difficult and results in strong compression artifacts, frame drops, and freezes. In the Low-bitrate video mode, the system remains pixel-streaming, but reduces $R_{\mathrm{vid}}(t)$ by lowering spatial resolution and/or frame rate. A practical engineering model is that video bitrate scales approximately with the number of pixels encoded per second. Let the spatial scale factor be $\eta\in(0,1]$ such that the width and height are scaled by $\eta$, and let the temporal scale factor be $\gamma\in(0,1]$ such that the frame rate is scaled by $\gamma$. Since the number of pixels per frame scales with area, the area scales as $\eta^2$. Hence the reduced video bitrate can be approximated by equation~\ref{eq:lb_video_scaling}
\begin{equation}
\label{eq:lb_video_scaling}
R_{\mathrm{vid}}^{(\mathrm{LB})}(t)\approx \eta^{2}\gamma \, R_{\mathrm{vid}}^{(\mathrm{Normal})}(t),
\end{equation}
and therefore
\begin{equation}
\label{eq:lb_mode_total}
R_{\mathrm{tot}}^{(\mathrm{LB})}(t)\approx R_{\mathrm{aud}}(t)+R_{\mathrm{vid}}^{(\mathrm{LB})}(t).
\end{equation}
Equation~\eqref{eq:lb_video_scaling} expresses an intuitive idea: halving both width and height reduces pixels by roughly a factor of four, and reducing frame rate reduces pixels per second further. This mode aims to delay the point at which pixel streaming becomes unusable, but it still depends on sending video frames continuously. In the AI reconstruction mode, the system stops transmitting pixel video and instead transmits a reconstructible representation. The transmitted stream is dominated by speech audio plus lightweight control signals and occasional reference updates in equation~\ref{eq:ai_mode_total}
\begin{equation}
\label{eq:ai_mode_total}
R_{\mathrm{tot}}^{(\mathrm{AI})}(t)\approx R_{\mathrm{aud}}(t)+R_{\mathrm{ctrl}}(t)+R_{\mathrm{ref}}(t),
\qquad
R_{\mathrm{vid}}(t)\approx 0.
\end{equation}
The important property is that $R_{\mathrm{ctrl}}(t)$ and $R_{\mathrm{ref}}(t)$ can be designed to be small compared to pixel video. This makes AI mode feasible in regimes where even aggressively compressed video is unstable. A compact control stream can be modeled as follows. Suppose the control signal is sent at a rate of $f_{\mathrm{ctrl}}$ control frames per second. Each control frame contains $N$ motion descriptors (for example, landmark points), and each descriptor is encoded using $d$ bits after quantization/packing. Then the control bitrate is approximately
\begin{equation}
\label{eq:ctrl_rate}
R_{\mathrm{ctrl}}\approx \frac{f_{\mathrm{ctrl}}\,N\,d}{1000}.
\end{equation}
All terms in \eqref{eq:ctrl_rate} have a direct physical meaning: $f_{\mathrm{ctrl}}$ is how often control information is sent, $N$ is how many descriptors are sent per control frame, and $d$ is the number of bits used for each descriptor. The division by $1000$ converts bits/s to kbps. This formula also makes the main design trade-offs clear: increasing $f_{\mathrm{ctrl}}$ can improve temporal smoothness, increasing $N$ can improve geometric detail, and increasing $d$ reduces quantization error; all three increase bitrate. Reference updates are occasional and amortized over time. Let each reference update be $S_{\mathrm{ref}}$ bits and let the time between updates be $T_{\mathrm{ref}}$ seconds. The average reference bitrate is
\begin{equation}
\label{eq:ref_rate}
R_{\mathrm{ref}}\approx \frac{S_{\mathrm{ref}}}{1000\,T_{\mathrm{ref}}}.
\end{equation}
Equation~\eqref{eq:ref_rate} is an average-rate calculation: a larger reference or a more frequent update increases $R_{\mathrm{ref}}$, while infrequent updates keep it small. Mode selection is performed using the smoothed goodput $\tilde{G}(t)$ as a proxy for available capacity and stability. Two thresholds $\tau_{\text{low}}$ and $\tau_{\text{high}}$ are used to create hysteresis, with $\tau_{\text{low}}<\tau_{\text{high}}$. Let the mode be $m(t)\in\{\mathrm{Normal},\mathrm{LB},\mathrm{AI}\}$. A simple hysteresis rule is:
\begin{equation}
\label{eq:hysteresis_rule}
m(t) =
\begin{cases}
\text{AI (e.g., Wav2Lip) (BR3)}, 
  & \tilde{G}(t) < \tau_{\text{low}},\\[4pt]
\text{Normal WebRTC (e.g., VP8) (BR1)}, 
  & \tilde{G}(t) \ge \tau_{\text{high}},\\[4pt]
\text{Low Bitrate (e.g., AV1) (BR2)}, 
  & \tau_{\text{low}} \le \tilde{G}(t) < \tau_{\text{high}}.
\end{cases}
\end{equation}

The role of hysteresis is to prevent rapid oscillations when the network fluctuates around a single threshold. In plain terms, the system only upgrades back to Normal after the network has clearly improved above $\tau_{\text{high}}$, and it only downgrades to AI after the network has clearly fallen below $\tau_{\text{low}}$. A necessary feasibility condition is that the chosen mode must fit within the estimated capacity. Using $\tilde{G}(t)$ as a capacity proxy, feasibility can be expressed as
\begin{equation}
\label{eq:feasibility}
R_{\mathrm{tot}}^{(m)}(t)\le \tilde{G}(t),
\end{equation}
where $R_{\mathrm{tot}}^{(m)}(t)$ is computed from \eqref{eq:normal_mode}, \eqref{eq:lb_mode_total}--\eqref{eq:lb_video_scaling}, or \eqref{eq:ai_mode_total} depending on the selected mode. If \eqref{eq:feasibility} is violated in Normal mode, the controller should reduce video load (LB) or replace video with reconstruction (AI). If \eqref{eq:feasibility} is violated even in AI mode, the remaining feasible fallback is audio-only because audio typically has the smallest and most stable bitrate term in \eqref{eq:total_rate}. The AI reconstruction path changes the meaning of the transmitted video modality while preserving the interactive nature of the session. Audio remains real-time and drives the generated mouth motion; the visual stream becomes a rendered output synthesized from the reference and speech. This design is especially useful in severe bandwidth regimes because it avoids continuously sending pixel video, which is the dominant term in \eqref{eq:normal_mode} and the hardest term to preserve under constrained links. The mathematical rate decomposition \eqref{eq:total_rate}--\eqref{eq:ai_mode_total} makes the core benefit explicit: AI mode removes $R_{\mathrm{vid}}(t)$ and replaces it with the controllable terms $R_{\mathrm{ctrl}}(t)$ and $R_{\mathrm{ref}}(t)$, whose average rates can be set by design using \eqref{eq:ctrl_rate} and \eqref{eq:ref_rate}. The telemetry-based capacity estimate \eqref{eq:tx_rate}--\eqref{eq:ewma} and the switching rule \eqref{eq:hysteresis_rule} provide a simple and stable mechanism to move between modes as network conditions change.

\section{Adaptive Bandwidth Controller}

The adaptive bandwidth controller operates in a closed loop that repeatedly estimates a stable notion of usable network capacity from real-time WebRTC telemetry and then selects an operating mode (Normal, Low-bitrate, or AI reconstruction) using hysteresis so that the call does not rapidly oscillate when the link fluctuates. The controller runs at discrete instants $t_k=k\Delta t$, where $\Delta t$ denotes the telemetry sampling interval. The instantaneous uplink and downlink throughputs, denoted by $\hat{R}_{\mathrm{tx}}(t_k)$ and $\hat{R}_{\mathrm{rx}}(t_k)$, are computed from cumulative byte counters observed at the endpoint. In equation~\ref{eq:ctr_tx}, $B_{\mathrm{tx}}(t_k)$ and $B_{\mathrm{rx}}(t_k)$ represent the cumulative transmitted and received bytes measured up to time $t_k$, while the differences $B_{\mathrm{tx}}(t_k)-B_{\mathrm{tx}}(t_{k-1})$ and $B_{\mathrm{rx}}(t_k)-B_{\mathrm{rx}}(t_{k-1})$ represent the bytes transferred during the most recent interval. The factor of $8$ in equation~\ref{eq:ctr_tx} converts bytes into bits, and division by $1000\Delta t$ yields throughput in kbps:
\begin{equation}
\label{eq:ctr_tx}
\hat{R}_{\mathrm{tx}}(t_k)=\frac{8\big(B_{\mathrm{tx}}(t_k)-B_{\mathrm{tx}}(t_{k-1})\big)}{1000\,\Delta t},\qquad
\hat{R}_{\mathrm{rx}}(t_k)=\frac{8\big(B_{\mathrm{rx}}(t_k)-B_{\mathrm{rx}}(t_{k-1})\big)}{1000\,\Delta t}.
\end{equation}
Because the raw estimates in equation~\ref{eq:ctr_tx} can vary due to short-term queueing, scheduling, and bursty packetization, the controller uses exponential smoothing to obtain more stable values. In equation~\ref{eq:ctr_ewma_rate}, $\tilde{R}_{\mathrm{tx}}(t_k)$ and $\tilde{R}_{\mathrm{rx}}(t_k)$ are smoothed throughputs, and $\alpha\in(0,1)$ controls the trade-off between responsiveness and stability, where a larger $\alpha$ retains more history and reduces sensitivity to spikes:
\begin{equation}
\label{eq:ctr_ewma_rate}
\tilde{R}_{\mathrm{tx}}(t_k)=\alpha\,\tilde{R}_{\mathrm{tx}}(t_{k-1})+(1-\alpha)\hat{R}_{\mathrm{tx}}(t_k),\qquad
\tilde{R}_{\mathrm{rx}}(t_k)=\alpha\,\tilde{R}_{\mathrm{rx}}(t_{k-1})+(1-\alpha)\hat{R}_{\mathrm{rx}}(t_k).
\end{equation}

The controller also incorporates packet loss because loss directly affects decodability, freeze events, and the quality of both audio and video. Packet loss over the most recent interval is estimated using cumulative RTP counters. In equation~\ref{eq:ctr_loss}, $P_{\mathrm{lost}}(t_k)$ denotes the cumulative number of lost packets and $P_{\mathrm{recv}}(t_k)$ denotes the cumulative number of received packets. The operator $\Delta$ in equation~\ref{eq:ctr_loss} indicates an interval difference such that $\Delta P(t_k)=P(t_k)-P(t_{k-1})$, and the ratio measures the fraction of packets that were lost during the last telemetry window:
\begin{equation}
\label{eq:ctr_loss}
\hat{p}_{\mathrm{loss}}(t_k)=\frac{\Delta P_{\mathrm{lost}}(t_k)}{\Delta P_{\mathrm{lost}}(t_k)+\Delta P_{\mathrm{recv}}(t_k)},
\qquad
\Delta P(t_k)=P(t_k)-P(t_{k-1}).
\end{equation}
A useful single-number proxy for how much of the transmitted bitrate is expected to arrive successfully is the effective goodput. As defined in equation~\ref{eq:ctr_goodput}, $\tilde{G}(t_k)$ discounts the smoothed uplink throughput by the interval loss ratio, so that higher packet loss reduces the estimated usable budget even if the sender is transmitting at the same rate:
\begin{equation}
\label{eq:ctr_goodput}
\tilde{G}(t_k)=\tilde{R}_{\mathrm{tx}}(t_k)\big(1-\hat{p}_{\mathrm{loss}}(t_k)\big).
\end{equation}
Interactive conferencing quality is additionally impacted by delay and delay variation. A conservative capacity proxy can therefore penalize $\tilde{G}(t_k)$ using measured round-trip time and jitter. In equation~\ref{eq:ctr_capacity}, $\widehat{RTT}(t_k)$ represents the measured round-trip time and $\hat{J}(t_k)$ represents measured jitter, while $\kappa$ and $\lambda$ are nonnegative weights that determine how strongly these terms reduce the effective capacity:
\begin{equation}
\label{eq:ctr_capacity}
\tilde{C}(t_k)=\frac{\tilde{G}(t_k)}{1+\kappa\,\widehat{RTT}(t_k)+\lambda\,\hat{J}(t_k)}.
\end{equation}
The quantity $\tilde{C}(t_k)$ in equation~\ref{eq:ctr_capacity} acts as the controller’s stability-aware estimate of what the network can support at time $t_k$.

Once the controller has $\tilde{C}(t_k)$, it selects the operating mode using hysteresis thresholds. The thresholds are denoted by $\tau_{\text{low}}$ and $\tau_{\text{high}}$ with $\tau_{\text{low}}<\tau_{\text{high}}$, and hysteresis is enforced by requiring that a threshold condition persist for a minimum duration before changing modes. This persistence is implemented through counters. In equation~\ref{eq:ctr_stable}, $n_{\downarrow}(t_k)$ counts consecutive samples in which $\tilde{C}(t_k)$ stays below $\tau_{\text{low}}$, and $n_{\uparrow}(t_k)$ counts consecutive samples in which $\tilde{C}(t_k)$ stays at or above $\tau_{\text{high}}$. The indicator function $\mathbb{I}[\cdot]$ equals $1$ when its condition is true and equals $0$ otherwise, which ensures that each counter resets when the corresponding condition fails:
\begin{equation}
\label{eq:ctr_stable}
n_{\downarrow}(t_k)=\mathbb{I}[\tilde{C}(t_k)<\tau_{\text{low}}]\,(n_{\downarrow}(t_{k-1})+1),\qquad
n_{\uparrow}(t_k)=\mathbb{I}[\tilde{C}(t_k)\ge\tau_{\text{high}}]\,(n_{\uparrow}(t_{k-1})+1).
\end{equation}
A minimum stability duration $T_{\text{stable}}$ corresponds to a minimum number of consecutive samples $N_{\text{stable}}=\lceil T_{\text{stable}}/\Delta t\rceil$. The mode decision rule in equation~\ref{eq:ctr_mode} uses these counters to prevent rapid toggling. The mode variable $m(t_k)$ takes values in the set $\{\mathrm{Normal},\mathrm{LB},\mathrm{AI}\}$, where $\mathrm{LB}$ denotes low-bitrate pixel streaming. As mentioned in equation~\ref{eq:ctr_mode}, the controller moves to AI reconstruction only after the network has remained below the lower threshold long enough, returns to Normal only after the network has remained above the upper threshold long enough, and otherwise stays in the intermediate Low-bitrate mode:
\begin{equation}
\label{eq:ctr_mode}
m(t_k)=
\begin{cases}
\mathrm{AI}, & n_{\downarrow}(t_k)\ge N_{\text{stable}},\\
\mathrm{Normal}, & n_{\uparrow}(t_k)\ge N_{\text{stable}},\\
\mathrm{LB}, & \text{otherwise}.
\end{cases}
\end{equation}

After selecting a mode, the controller ensures that the expected transmission demand fits within the estimated capacity. The total outgoing bitrate demand is decomposed into components in equation~\ref{eq:ctr_total}. In that equation, $R_{\mathrm{aud}}(t_k)$ represents the audio bitrate, $R_{\mathrm{vid}}(t_k)$ represents the encoded pixel-video bitrate, $R_{\mathrm{ctrl}}(t_k)$ represents the control-stream bitrate used for reconstruction, and $R_{\mathrm{ref}}(t_k)$ represents the amortized bitrate of occasional reference updates:
\begin{equation}
\label{eq:ctr_total}
R_{\mathrm{tot}}(t_k)=R_{\mathrm{aud}}(t_k)+R_{\mathrm{vid}}(t_k)+R_{\mathrm{ctrl}}(t_k)+R_{\mathrm{ref}}(t_k).
\end{equation}
The feasibility condition is expressed in equation~\ref{eq:ctr_feasible}, where $R_{\mathrm{tot}}^{(m)}(t_k)$ denotes the demand implied by the selected mode $m(t_k)$. The inequality in equation~\ref{eq:ctr_feasible} states that the chosen mode should not demand a sustained bitrate larger than what the link can stably deliver:
\begin{equation}
\label{eq:ctr_feasible}
R_{\mathrm{tot}}^{(m)}(t_k)\le \tilde{C}(t_k).
\end{equation}
The expression $R_{\mathrm{tot}}^{(m)}(t_k)$ is obtained from equation~\ref{eq:ctr_total} by noting which terms are active in each mode. In Normal mode, the dominant terms are $R_{\mathrm{aud}}(t_k)$ and $R_{\mathrm{vid}}(t_k)$, so the demand is approximately $R_{\mathrm{aud}}(t_k)+R_{\mathrm{vid}}(t_k)$. In AI mode, the pixel-video term is suppressed, so the demand is approximately $R_{\mathrm{aud}}(t_k)+R_{\mathrm{ctrl}}(t_k)+R_{\mathrm{ref}}(t_k)$. In Low-bitrate mode, the controller reduces pixel-video demand by scaling down resolution and/or frame rate. A simple engineering approximation for how video bitrate changes under spatial and temporal scaling is given in equation~\ref{eq:ctr_scale}. In that equation, $\eta\in(0,1]$ is the spatial scale factor applied to both width and height, and $\gamma\in(0,1]$ is the temporal scale factor applied to frame rate. The factor $\eta^2$ reflects that pixel count per frame scales with image area, and the factor $\gamma$ reflects that frames per second are reduced:
\begin{equation}
\label{eq:ctr_scale}
R_{\mathrm{vid}}^{(\mathrm{LB})}(t_k)\approx \eta^2\gamma\,R_{\mathrm{vid}}^{(\mathrm{Normal})}(t_k).
\end{equation}
As a result, the Low-bitrate demand becomes approximately $R_{\mathrm{aud}}(t_k)+R_{\mathrm{vid}}^{(\mathrm{LB})}(t_k)$, and the controller can choose $(\eta,\gamma)$ so that equation~\ref{eq:ctr_feasible} is satisfied.

In AI reconstruction mode, the bitrate contributions that replace pixel video are explicitly controllable. The control-stream bitrate can be modeled by the update rate and payload size. As defined in equation~\ref{eq:ctr_ctrl_rate}, $f_{\mathrm{ctrl}}$ denotes the number of control updates per second, $N$ denotes the number of descriptors sent per update, and $d$ denotes the number of bits used to encode each descriptor after quantization/packing. Dividing by $1000$ converts bits/s to kbps:
\begin{equation}
\label{eq:ctr_ctrl_rate}
R_{\mathrm{ctrl}}\approx \frac{f_{\mathrm{ctrl}}\,N\,d}{1000}.
\end{equation}
Similarly, the reference-update contribution is the amortized cost of infrequent identity refreshes. As mentioned in equation~\ref{eq:ctr_ref_rate}, $S_{\mathrm{ref}}$ is the size of one reference update in bits and $T_{\mathrm{ref}}$ is the time between two updates in seconds:
\begin{equation}
\label{eq:ctr_ref_rate}
R_{\mathrm{ref}}\approx \frac{S_{\mathrm{ref}}}{1000\,T_{\mathrm{ref}}}.
\end{equation}
Equations~\ref{eq:ctr_ctrl_rate} and \ref{eq:ctr_ref_rate} make the controller knobs explicit. When $\tilde{C}(t_k)$ decreases, the controller can reduce $f_{\mathrm{ctrl}}$, reduce $N$, reduce $d$ by using coarser quantization, and increase $T_{\mathrm{ref}}$ so that equation~\ref{eq:ctr_feasible} remains satisfied while keeping $R_{\mathrm{aud}}(t_k)$ real-time. When $\tilde{C}(t_k)$ increases, the controller can reverse these adjustments to improve visual stability and identity retention.

Overall, the controller is driven by measurable quantities and simple, auditable rules. The throughput estimates in equation~\ref{eq:ctr_tx}, the smoothing in equation~\ref{eq:ctr_ewma_rate}, the loss and goodput calculations in equations~\ref{eq:ctr_loss} and \ref{eq:ctr_goodput}, and the stability-aware capacity proxy in equation~\ref{eq:ctr_capacity} produce a real-time estimate of what the link can sustain. The hysteresis and persistence logic in equations~\ref{eq:ctr_stable} and \ref{eq:ctr_mode} prevents unstable switching. The feasibility constraint in equation~\ref{eq:ctr_feasible} ties the selected operating mode to the measured capacity. The key benefit is that, when severe bandwidth constraints make $R_{\mathrm{vid}}(t_k)$ infeasible, switching to AI reconstruction removes the dominant pixel-video demand term from equation~\ref{eq:ctr_total} and replaces it with the compact, design-controlled terms $R_{\mathrm{ctrl}}$ and $R_{\mathrm{ref}}$ governed by equations~\ref{eq:ctr_ctrl_rate} and \ref{eq:ctr_ref_rate}, thereby maintaining conversational continuity under extremely constrained network conditions. All studies are conducted with Wav2Lip as the talking-head backend. We selected Wav2Lip due to its widespread adoption, open-source nature, and the fact that it establishes a robust foundation for lip-sync quality. However, the system is completely compatible with alternative THG backends (MakeItTalk, SadTalker, and VadicTHG), which we have integrated behind the same inference API. It is important to note that the controller, SFU, and WebRTC pipeline are not impacted by the backend switch.

\section{Experimental Setup}

This section shows how the system is assessed in a method that is repeatable, fair to all baselines, and closely linked with the aim of preserving interactive communication under severe bandwidth restrictions.  The assessment is structured on four ideas:  (i) using the same conferencing task and the same test content across methods, (ii) controlling the network so that bandwidth, loss, delay, and jitter can be varied in a known way, (iii) measuring both network-level quality of service (QoS) and user-facing quality of experience (QoE), and (iv) separating offline model quality from end-to-end conferencing behavior.  Throughout the tests, all algorithms work in real time with the same call length and the same audio content, and all measurements are obtained at a set sampling interval $\Delta t$ so that time-aligned comparisons can be performed. 

The experimental system comprises of two endpoints (caller and callee) and, when required, a forwarding server that operates as an SFU.  Each endpoint is a normal WebRTC participant that takes microphone audio and (when video is enabled) camera video.  The signaling channel is used exclusively to create sessions; all material is delivered across real-time secure channels.  The studies include both peer-to-peer calls and SFU-routed calls since multi-party systems commonly deploy an SFU even for two-party connections, and because SFU forwarding may modify loss patterns and delay profiles.  For every configuration, the same test procedure is followed: the call is established, a short warm-up period is allowed for encoder and congestion-control stabilization, the network condition is applied, and then the system runs for a fixed evaluation window during which all telemetry and quality measurements are recorded. 

Two sorts of evaluation inputs are employed.  The first kind is a live interactive call, where participants converse spontaneously.  This option captures genuine conversational dynamics like as pauses, turn-taking, and fluctuating volume, and it helps test that the system operates appropriately under real use.  The second kind is a controlled playback configuration, where a pre-recorded audio track (and optionally a reference face video) is utilized as the input.  This second parameter is significant because it makes the tests repeatable: every method gets precisely the same speech material in the same time, which prevents confusing effects from variable speaking styles between runs.  When a controlled playback is utilized, the same reference face picture is also used for all AI reconstructions so that identity criteria remain constant. 

The experiments are aimed to examine the complete bandwidth range relevant to the study, with specific attention on the severe regime.  The network is regulated via traffic shaping so that the available bandwidth, latency, jitter, and packet loss may be configured directly.  Bandwidth control is enforced as a maximum rate on the sender’s uplink since uplink is frequently the principal bottleneck in video conferencing.  Let $B_{\max}$ be the specified bandwidth limit for a run.  In each run, the limiter mandates that the transmitted data rate does not exceed $B_{\max}$ across brief windows, which pushes the encoder and congestion control to adjust.  In addition to bandwidth limits, separate impairments are implemented to produce genuine tough situations.  A fixed one-way delay may be provided to mimic lengthy pathways, and random jitter can be added to model variable queueing.  Packet loss may be implemented either as uniform random loss or as burst loss, where bursts are more indicative of wireless interference and busy queues.  These controlled impairments enable the studies to include both ``clean low bandwidth'' and ``low bandwidth with instability,'' since actual networks typically contain both.  Each impairment profile is performed numerous times so that variability may be noticed, and the identical profiles are applied to every baseline so that comparisons are fair. 

All baselines are assessed under precisely the same network circumstances and with the same input content.  The first baseline is ordinary WebRTC video conferencing in which both voice and pixel video are delivered using traditional encoding and transport.  This baseline illustrates how normal browser-based conferencing works when it depends on codec rate control and congestion management to endure bandwidth decreases.  The second baseline is an audio-only mode, in which video is deactivated and only microphone audio is conveyed.  This baseline is significant since audio-only is typically the backup utilized by actual goods, and it gives a lower constraint on bandwidth utilization and a reference for conversational continuity without visual signals.  The third baseline is low-bitrate pixel video, in which resolution and/or frame rate are actively decreased to try to keep within the bandwidth allowance while still delivering video frames.  This baseline differentiates between ``aggressive compression of pixels'' and ``semantic replacement by reconstruction,'' which is the core contribution of this study.  The proposed system is assessed in its full adaptive form, where it picks modes based on telemetry, and also in fixed-mode settings, where the system is compelled to stay in one mode for the complete run.  Fixed-mode runs are utilized to identify how much of the performance comes from the AI reconstruction itself vs the switching logic. 

A significant aspect of the experimental setup is the logging and measurement pipeline.  The endpoints capture time-resolved telemetry at interval $\Delta t$, generating synchronized time series of throughput, loss, jitter, round-trip time, and video rendering information.  Let $t_k=k\Delta t$ be the sampling time.  The uplink and downlink throughputs are calculated as in equation~\ref{eq:ctr_tx}, using cumulative byte counters.  Packet loss is calculated as in equation~\ref{eq:ctr_loss}.  These metrics are recorded in a row at each $t_k$, which enables charts such as throughput-versus-time and loss-versus-time to be constructed.  In addition to transport counters, the endpoints additionally track media-level information such as the transmitted frame rate, the effective resolution, and any times when video freezes.  A freeze event is recognized when the displayed frame timestamp does not progress for longer than a preset freeze threshold $T_{\mathrm{freeze}}$ while the call is still running.  The overall freeze time for a run is thus the sum of all freeze durations, and the freeze ratio is determined as the proportion of the evaluation window spent frozen.  This freeze ratio is a straightforward and interpretable indicator of how often the viewer encounters ``stuck video.''  For audio, dropouts are detected by watching gaps in the playback buffer or by noticing discontinuities in RTP audio timestamps; the entire audio dropout duration is logged similarly. 

For the proposed AI reconstruction mode, extra instrumentation is incorporated to quantify end-to-end generation latency and its contribution to conversational delay.  Let $t_k^{\mathrm{cap}}$ be the time when a reference frame and an audio segment are captured, let $t_k^{\mathrm{req}}$ be the time when the inference request is sent, let $t_k^{\mathrm{resp}}$ be the time when the synthesized segment is received, and let $t_k^{\mathrm{play}}$ be the time when it is first rendered as part of the outgoing video track.  The synthesis pipeline latency may be split into (i) capture/packaging delay, (ii) network request/response delay, (iii) inference compute time, and (iv) decoding/render delay.  The end-to-end synthesis delay is defined as in equation~\ref{eq:synth_latency}.
\begin{equation}
\label{eq:synth_latency}
T_{\mathrm{synth}}(k)=t_k^{\mathrm{play}}-t_k^{\mathrm{cap}}.
\end{equation}
Logging the components that constitute $T_{\mathrm{synth}}(k)$ makes it evident whether delays are dominated by compute or by transport, and it enables the system architecture to be modified for interactive usage.  In addition, mode-switching delay is assessed.  If the controller chooses to switch at time $t_k$, the mode-switch completion time is the first moment when the outgoing track is entirely in the new mode.  The switching latency is the difference between those two times.  This measurement is critical because frequent switching is harmful to users even if average QoE is good. Evaluation measures are designed to represent both efficiency and user enjoyment.  Network efficiency is measured by average and percentile throughput as well as total data utilization throughout the assessment timeframe.  Let $\hat{R}_{\mathrm{tx}}(t_k)$ be the instantaneous uplink throughput from equation~\ref{eq:ctr_tx}.  The average uplink throughput for a run of $K$ samples is determined as in equation~\ref{eq:avg_tx}.
\begin{equation}
\label{eq:avg_tx}
\overline{R}_{\mathrm{tx}}=\frac{1}{K}\sum_{k=1}^{K}\hat{R}_{\mathrm{tx}}(t_k),
\end{equation}
The total transferred data is determined by translating the cumulative byte counter difference throughout the run into megabytes.  Stability is measured by packet loss ratio, jitter, and freeze ratio.  Responsiveness is measured by end-to-end latency and by the time necessary to recover from a bandwidth loss.  Recovery time is defined as the time from a dramatic decrease in $\tilde{C}(t_k)$ to the time when the selected mode becomes viable again according to equation~\ref{eq:ctr_feasible}, implying the call recovers to a stable steady state.  For video quality, two groups of measures are examined.  The first class comprises of typical perceptual and structural metrics calculated on decoded pixel frames when pixel video is available.  The second class comprises of reconstruction-specific criteria such as identity consistency and lip-sync consistency between audio and produced mouth movements.  Lip-sync may be assessed by assessing alignment between audio characteristics and mouth-region movements, and identity can be evaluated by comparing face embeddings between the reference identity and the produced frames.  These metrics are calculated on the output received by the far end, since that output is what the user really sees. To guarantee that comparisons are fair, every run follows the same control methodology.  The same bandwidth profile is applied to all methods, the same length and warm-up are utilized, and all logging is conducted at the same interval.  Randomized impairments such as loss are created with fixed seeds such that the same loss pattern may be repeated for each approach.  Each scenario is replicated numerous times, and the presented findings reflect both central tendency and variability.  Where subjective quality is assessed, human evaluations are obtained under a consistent viewing and hearing arrangement and with randomized presentation order to prevent bias.  Finally, the experiments split two deployment conditions: one where inference is conducted on a local computer (representing on-device or edge inference) and one where inference is performed on a distant server (representing cloud inference).  This division enables the study to show how synthesis delay and network overhead fluctuate depending on where the AI module is situated, while keeping the essential bandwidth-saving behavior of AI mode constant. Overall, the experimental setting is aimed to answer three questions clearly.  First, under severe bandwidth limits, can the system preserve conversational continuity better than pixel-video baselines, as shown in freeze ratio, audio stability, and mode feasibility.  Second, can the AI reconstruction mode give a useful video-like experience, as evidenced in lip-sync and identity consistency metrics calculated on the received output.  Third, can the adaptive controller transition modes reliably and with little disturbance, as shown in switching delay and the lack of oscillation under varying bandwidth.  The remainder of the article provides outcomes utilizing these controlled circumstances and the related documented data.

\section{Results}

\begin{figure}[t]
\centering
\includegraphics[width=0.5\linewidth]{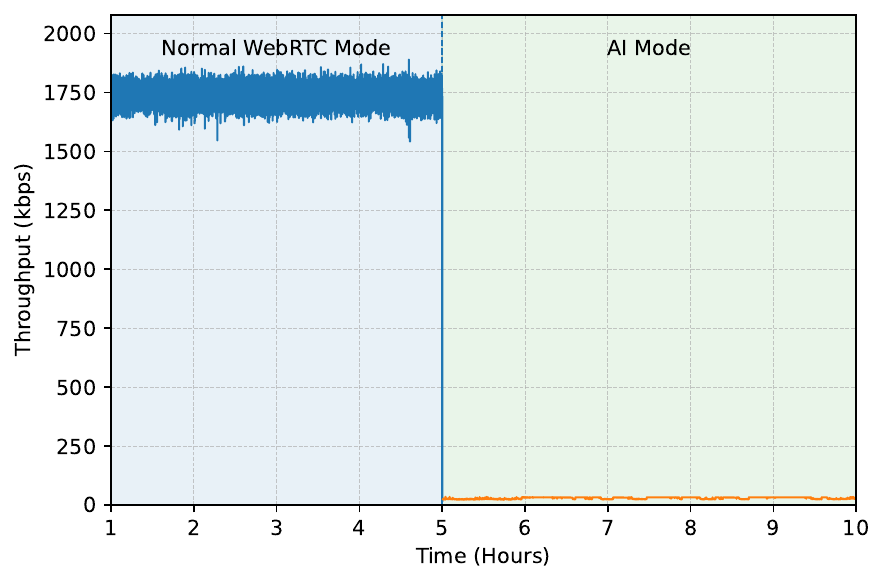}
\caption{Comparative analysis of throughput (kbps) for Standard WebRTC vs. AI Synthesis. Note the significant bandwidth saving achieved during the AI Synthesis phase compared to the Standard WebRTC baseline.}
\label{fig:20h_trace}
\end{figure}

This section illustrates the observed behavior of the conferencing system utilizing (i) time-resolved throughput traces, (ii) long-run bandwidth logs with a clear mode-change event, and (iii) summary tables that condense the time series into interpretable data.  The purpose of this investigation is to explain, in a transparent and repeatable manner, how the system acts while it runs as traditional WebRTC video and how that behavior changes after switching to the AI reconstruction mode.  All charts and tables utilize the same fundamental values taken from the recorded counters.  The transmit and receive throughputs depicted in Figure~\ref{fig:20h_trace} are estimated from cumulative byte counters using the usual differencing approach described in equation~\ref{eq:ctr_tx}.  In the same manner, the long-run uplink trace in Figure~\ref{fig:20h_trace} employs the recorded uplink throughput $\hat{R}_{\mathrm{tx}}(t)$ aggregated at a given sampling interval, which is the major indication of how much uplink capacity the session consumes over time.  Figure~\ref{fig:20h_trace} shows a fine-grained perspective of throughput progression throughout a single session.  The horizontal axis indicates time in seconds while the vertical axis shows throughput in kbps.  The two curves correspond to the sender-side throughput and the receiver-side throughput, represented in the graphic as ``Send kbps'' and ``Receive kbps''.  The send curve approximates the instantaneous outgoing application traffic (including RTP overheads) and hence depicts the demand imposed on the uplink.  The receive curve approximates the instantaneous receiving traffic at the peer, which is often somewhat lower than the transmit curve owing to packet loss, retransmission behavior, and measurement asymmetries.  When the session is stable, both curves should grow smoothly and stay near to each other, showing that the network is transmitting the specified media rate without severe loss or congestion.  When the network gets restricted, the transmit throughput tends to display adaption patterns such as step-downs or oscillations, reflecting the combined behavior of encoder rate control and congestion management.  At the same time, the receive throughput may drift downward more rapidly if loss rises, since fewer packets are transmitted to the receiver.  Therefore, the gap between the two curves is itself an indirect sign of instability: a bigger persistent gap typically suggests more loss or more aggressive falling along the course, whereas a near-overlap indicates good delivery.  In the context of this paper, Figure~\ref{fig:20h_trace} is used mainly to validate that the logging pipeline correctly captures time-resolved send/receive bandwidth and to illustrate how quickly the system reacts to changing conditions at sub-minute scales, which is essential for any real-time controller.  The interpretation of Figure~\ref{fig:20h_trace} is likewise closely connected to the controller quantities.  As noted in equation~\ref{eq:ctr_tx}, throughput at time $t_k$ is determined by differencing cumulative bytes across $\Delta t$; as a consequence, short-term bursts within an interval appear as local oscillations while the overall trend represents the effective sending rate.  When the controller utilizes a smoothed capacity proxy such as $\tilde{C}(t_k)$ in equation~\ref{eq:ctr_capacity}, the objective is to avoid responding to these minor swings and instead respond to persistent changes.  Therefore, the throughput-over-time plot gives a visual reason for implementing smoothing and persistence in the controller: it indicates that even under steady operation the measured rate is not exactly constant, and it displays the time scale at which changes occur.  Figure~\ref{fig:20h_trace} is also helpful for testing if a mode transition produces noticeable disruption.  If a switch is implemented correctly, the plot should show a clean transition in send throughput from a higher regime (pixel video) to a lower regime (reconstruction overhead), without long periods of zero throughput (which would indicate a stall) and without repeated toggling between levels (which would indicate unstable switching).  Thus, this image functions as an initial diagnostic and aids the eventual long-run research.  Figure~\ref{fig:20h_trace} exhibits a long-run uplink bandwidth trace covering an extended session in which the system operates in Normal pixel-video mode for the first half of the experiment and then transitions to AI reconstruction mode at a specified time.  The graphic includes a single main curve depicting the uplink throughput and a vertical dashed marker marking the point at which AI mode is activated.  This figure has two uses.  First, it gives a clear visual comparison of the operating regimes:  the regular video mode often needs much greater uplink capacity since the sent payload comprises compressed pixel frames in addition to audio.  Second, it reveals that the decrease following the AI transition is not a short-lived artifact but a stable condition over a long duration, which is essential since brief testing might disguise instability, drift, or steady rate rises. 
The dashed line in Figure~\ref{fig:20h_trace} is a structural aspect of the study since it anchors a before/after comparison.  The region preceding the dotted line corresponds to the distribution of uplink values in Normal mode.  The region following the dotted line corresponds to the distribution under AI mode.  Because the logging is done every second, each segment has a significant number of data and so the comparison is statistically relevant even without training a model.  The major qualitative discovery is that the uplink curve drops abruptly at the changeover time and stays in a low, confined band afterward.  This behavior is precisely what the rate decomposition predicts.  In Normal mode, the total outgoing rate is dominated by the pixel-video term in the decomposition provided in equation~\ref{eq:ctr_total}.  In AI mode, the pixel-video term is suppressed and replaced by smaller control and reference terms as described in equation~\ref{eq:ai_mode_total}; consequently the predicted uplink demand lowers and stays low as long as audio and control signals are steady.  The long-run figure shows that this behavior is obtained in reality over several hours, which significantly supports the argument that the technique is acceptable for persistent conferencing in confined locations rather than merely for brief demonstrations. Table~\ref{tab:bandwidth_summary} reduces the long-run traces into summary data that enable straightforward comparison between modes.  The uplink mean and downlink mean quantify the average bandwidth utilized in each direction during the measurement frame.  The total mean reflects the sum of uplink and downlink means, which is a useful single figure when the overall network cost is of relevance.  The total median gives a robust assessment of the ``typical'' operating point, eliminating the effect of rare spikes.  The total $p95$ provides the 95th percentile and captures near-worst-case demand; this is crucial for provisioning and for determining if the approach creates bursts that might overwhelm weak connections.  The ``Data/hour'' and ``Data for 5h'' columns translate average throughput into total transmitted data volume, which is a realistic indicator for customers on metered or limited connections.  The table supports two separate findings.  First, Normal mode requires large bandwidth since it continually sends pixel video.  This fits with equation~\ref{eq:normal_mode}, where the dominating cost is the video term.  Second, AI mode dramatically decreases both uplink and downlink consumption because the pixel-video term is deleted and replaced by the audio and compact control representation as described in equation~\ref{eq:ai_mode_total}.  The decrease is reflected not just in the mean but also in the median and the $p95$, which implies that the reduction is permanent and not reliant on unusual low-rate periods.  Interpreting the columns together is key.  A technique that has a low mean but a high $p95$ may look efficient on average but still create spikes that cause jitter or buffering; in contrast, a method that maintains both median and high percentiles low is more likely to stay steady on weak connections.  In this regard, Table~\ref{tab:bandwidth_summary} presents proof that AI mode delivers not just a lower average rate but also a more predictable rate envelope over lengthy periods. Each row refers to a bandwidth limit and a technique.  The WebRTC video baseline reflects a typical pixel-streaming technique under the same limitation.  The audio-only baseline reflects the minimal-bandwidth fallback employed by many systems and offers a reference point for conversational continuity without visual presence.  The suggested technique, labeled as ``Ours (AI)'', reflects the semantic substitution approach, where visual presence is rebuilt from auditory and compact control signals rather than communicated as pixels.  The columns in Table~\ref{tab:bandwidth_summary} are selected to represent both user perception and system stability.  The QoE column describes perceived quality (for example, via MOS or a similarly specified score).  The lip-sync column indicates how effectively mouth movements matches with the audio.  The identification column evaluates how closely the rebuilt face matches the reference identity.  The freeze column evaluates video instability; lower is preferable since it shows fewer or shorter freezes.  The latency column measures end-to-end delay, where lower values facilitate natural turn-taking.  The structure of Table~\ref{tab:bandwidth_summary} is purposefully matched with the mode design.  Under severe caps, traditional WebRTC video typically becomes unstable since it still needs a minimum rate for video frames, and its visual stream is sensitive to loss and prediction-chain breakdown, which increases freezes.  Audio-only may preserve continuity but gives no visual presence, therefore lip-sync and identification are not relevant.  The suggested AI approach is likely to prevail in the severe regime since its transmitted payload is mostly audio and compact control, and so it is more practical under tight constraints while still giving a video-like output.  When supplied with measures, the table should indicate that ``Ours (AI)'' offers greater QoE and lower freeze than WebRTC video in the lowest caps, while delivering significant lip-sync and identification scores that audio-only cannot give.  For the paper narrative, the most essential interpretation of the table is across each cap with the best approach at 64--256 kbps should be the suggested way since this is exactly the regime in which semantic substitution is meant to beat intensive pixel compression. Each succeeding row eliminates or weakens one component, such as disconnecting the control signal, lowering reference update behavior, deleting temporal smoothing, or employing a smaller model.  The MOS/QoE column measures overall perceived quality, the lip-sync column catches audio-visual alignment, and the flicker column records temporal instability such as jittery facial animation, frame-to-frame inconsistencies, or identity drift.  The objective of this table is to prevent ambiguous conclusions.  If the whole system performs best but a smaller alternative works nearly as well, that suggests which elements may be simplified for deployment.  Conversely, if eliminating temporal smoothing produces a considerable rise in flicker, it illustrates that smoothing is not only decorative but needed for steady viewing.  Similarly, if deleting the control signal lowers lip-sync or identity stability, it supports the decision to send compact descriptors even in low-bandwidth mode.  For reviewers, ablations are particularly significant since they prove that performance is the product of the proposed system architecture rather than an unintended artifact of a single model configuration.  Mode change is only beneficial if it happens swiftly and without disturbing the discourse.  Handover delay is measured as the time between the moment the controller recognizes that a switching condition has been fulfilled and the moment the receiver sees the first stable frame generated by the new mode.  This description reflects the user-perceived idea of disruption since a changeover is only complete when the new visual output becomes steady at the receiving side.  In the controller, detection time corresponds to the first sample $t_k$ at which the persistence counters in equation~\ref{eq:ctr_stable} match the stability criterion and the mode choice in equation~\ref{eq:ctr_mode} changes state.  Completion time relates to the first moment when the outgoing track and rendering pipeline are generating frames in the new mode without gaps outside a short stability tolerance window.  The tests employ step adjustments in bandwidth using controlled traffic shaping such that threshold crossings are repeated.  Audio is kept continuous throughout, so that any disruption observed is primarily visual and attributable to the handover mechanism rather than to a complete session restart. In reporting changeover latency, it is necessary to include both central tendency and dispersion since a low mean with a broad tail might nevertheless induce occasional significant disturbances.  Therefore, the latency table should include mean and median values as well as measures of variability and extrema. The analysis should also separate transition types, since switching into AI mode can require initialization of reconstruction and control-stream handling, while switching back to pixel video can require encoder ramp-up. If the system avoids session renegotiation, then the latency should primarily reflect local pipeline reconfiguration and buffering. A successful handover mechanism should show bounded latencies and should not introduce audio dropouts, call drops, or repeated oscillation. When presented together with Figure~\ref{fig:20h_trace}, the handover analysis explains not only that bandwidth consumption drops after switching, but also that the drop can be achieved with minimal transient disturbance, which is essential for real-time conferencing.

\begin{table}[t]
\centering
\caption{Bandwidth and data usage}
\label{tab:bandwidth_summary}
\begin{tabular}{lrrrrrrr}
\toprule
\textbf{Mode} &
\textbf{Uplink} &
\textbf{Downlink} &
\textbf{Total} &
\textbf{Total} &
\textbf{Total} &
\textbf{Data/hour} &
\textbf{Data/5h} \\
&
\textbf{Mean} &
\textbf{Mean} &
\textbf{Mean} &
\textbf{Median} &
\textbf{p95} &
&
\\
&
(kbps) &
(kbps) &
(kbps) &
(kbps) &
(kbps) &
(GiB) &
(GiB) \\
\midrule
AV1 &
1486.33 & 1418.04 & 2904.38 & 2972.10 & 3503.70 & 1.31 & 12.76 \\
AI Mode &
28.23 & 15.42 & 43.65 & 32.80 & 64.70 & 0.02 & 0.19 \\
\bottomrule
\end{tabular}
\end{table}

\begin{table}[t]
\centering
\caption{Audio/Video jitter, RTT, and freeze count. Lower is better ($\downarrow$).}
\label{tab:av_quality_metrics}
\begin{tabular}{lrrrr}
\hline
Method & Audio Jitter (ms)$\downarrow$ & Video Jitter (ms)$\downarrow$ & RTT (ms)$\downarrow$ & Freeze$\downarrow$ \\
\hline
Mode~1 & 2.414 & 4.186 & 2.729 & 14 \\
Mode~2 & 2.138 & 5.090 & 18.910 & 17 \\
Mode~3  & 0.256 & 3.000 & 2.333 & 0 \\
\hline
\end{tabular}
\end{table}

\section{Discussion}

The findings indicate that replacing pixel video transmission with an audio-driven talking-head reconstruction approach may preserve a video-like conversational experience while running at very low bandwidth.  The key engineering cause for this trend is the shift in what dominates the transmitted payload.  In traditional conferencing, the encoded video term is often the greatest contribution to the overall bandwidth and is also the most vulnerable under congestion since inter-frame prediction makes the stream susceptible to packet loss and delay fluctuation.  When bandwidth diminishes, the encoder must drop quality and frame rate, and the transport layer may face greater loss, creating freezes and unsteady playback.  In the proposed technique, the AI reconstruction mode inhibits continuous pixel-video transmission and instead transmits real-time speech audio with compact control information and infrequent reference updates.  This shifts the bandwidth requirement from a high-variance, loss-sensitive pixel stream to a smaller and more controllable payload, which explains the large bandwidth reduction summarized in Table~\ref{tab:bandwidth_summary} . The throughput traces further suggest that the system may stay in a sustained low-rate regime once the AI mode is started, which is critical for practical implementation in networks that remain limited for extended periods rather than simply for brief bursts. Despite these benefits, the technique has severe limits that must be explicitly recognized.  First, the reconstructed video stream is not a faithful pixel-level depiction of the genuine camera feed; it is a synthetic approximation whose accuracy relies on the robustness of the talking-head model, the quality of the reference identity, and the features of the input audio.  If the reference picture is poorly lighted, partly obscured, or not indicative of the user’s usual look, identity consistency might suffer.  Similarly, if the speaker changes often or numerous voices are present, the reconstructed face may not match the current speaker unless the system contains speaker diarization and identity switching.  This may lead to an improper link between aural and visual identification, which may be inappropriate in particular circumstances such as formal meetings or clinical contacts.  Second, reconstruction quality might suffer for head positions, emotions, or occlusions not adequately represented by the model’s training distribution.  Extreme yaw angles, fast head movements, hand-to-face occlusions, microphones covering part of the lips, and low-light noise are prevalent in actual calls and may create artifacts, temporal flicker, or occasional frame instability.  Third, long-duration stability is challenging: generative outputs may drift slowly in appearance or display periodic artifacts depending on how the model handles temporal conditioning.  Even if average lip-sync is robust, occasional misalignment may occur when speech comprises quick phoneme transitions, background noise, or non-speech noises such as laughing, coughing, or overlapping speakers.  Fourth, AI mode creates compute reliance.  While bandwidth demand reduces, the endpoint or inference service must conduct model inference within real-time limitations.  If computation is inadequate or congestion develops, the system may exchange bandwidth savings for higher delay or lower frame rate.  This implies that for certain devices, particularly low-power endpoints, local inference may not be viable, and server or edge inference becomes essential, which brings extra network latency and potential privacy problems. Failure modes may be divided into transport-level, controller-level, and synthesis-level failures.  Transport-level failures include very high loss, significant jitter, or transient disconnections that impair real-time delivery; in such instances, even low-bitrate control signals may not arrive consistently, and the best attainable backup may be audio-only.  Controller-level errors include oscillation between modes when network circumstances linger near thresholds, delayed switching when smoothing is too slow, or premature switching when thresholds are too aggressive.  These dangers are lessened by hysteresis and persistence logic, but they remain susceptible to parameter selection and to the heterogeneity of actual networks.  Synthesis-level problems include faulty reference inputs, audio capture difficulties, model runtime faults, or unexpected material such as non-human sounds, which may yield undesirable video portions.  The system should consequently incorporate guardrails such as input validation, confidence estimate for landmark/control extraction, fallback to a static avatar or last-known-good frame, and quick reversion to audio-only when synthesis cannot achieve a minimal quality or latency constraint.  From a real-time perspective, one critical constraint is that switching should not require renegotiation of the underlying WebRTC session, because renegotiation introduces a large and unpredictable delay; maintaining a constant session while substituting the outgoing video track is therefore an important deployment property, but it also requires careful handling of timing and buffering so that receivers do not interpret the change as a stream failure. Privacy, ethics, and security are significant issues when synthetic media is brought into communication.  Even while AI mode may limit the quantity of raw video broadcast, it still incorporates biometric information and possibly sensitive material.  Reference photos and audio include individually identifiable clues, and reconstructed video is a synthetic portrayal of a real person.  This adds the potential of abuse, such as impersonation, false recordings, or the fabrication of misleading information.  Responsible deployment consequently demands unambiguous user permission and openness.  Users should be aware that their outgoing video is synthetic, and receivers should likewise be able to recognize that the stream is created, for example by a visual sign or metadata.  In addition, access control and data reduction should be enforced.  Reference pictures and audio segments should not be maintained longer than required for synthesis, and if server-side inference is employed, storage should be avoided by default or safeguarded under rigorous retention regulations.  From a transport standpoint, baseline confidentiality and integrity should be assured utilizing typical secure real-time media channels.  From an analytics standpoint, all telemetry collection must be linked with privacy standards; bandwidth and quality logs might show trends regarding user behavior and connection, therefore logs should be handled as sensitive operational data.  If the system communicates compact facial motion descriptors, it should be stated whether these descriptors may be used to reconstruct identity or expressions beyond what is intended; such representations may still be biometric, therefore their treatment should follow the same precautions as video. Deployment concerns include compatibility, scalability, latency, and operational resilience.  Compatibility is necessary because conferencing ecosystems are varied; not all clients support the same codecs, and not all environments enable the same network traversal behavior.  A realistic method is to guarantee that the AI mode can work without needing modifications on the receiver side beyond what a conventional WebRTC client can accept, for example by providing the synthesized video as a standard video track.  Scalability becomes important in multi-party sessions.  In pixel-video mode, an SFU may grow by forwarding streams, but in AI mode the system must select where synthesis happens.  If each sender synthesizes locally before transmitting, server cost is cheap but client compute needs grow.  If synthesis happens on an edge or cloud service, client needs reduce but server cost and network latency rise.  These trade-offs should be established depending on deployment restrictions, such as the availability of GPUs at the edge, estimated participant numbers, and legal considerations concerning where biometric processing is authorized.  Latency is crucial for conversational quality; consequently, the synthesis pipeline must be constructed such that the incremental delay of AI mode stays within acceptable interactive boundaries.  This entails careful selection of audio chunk sizes, buffering techniques, and frame generation cadence, as well as reducing the reference update cost.  Operational robustness involves monitoring and fallback systems.  The system should continually verify whether the current mode remains viable given observed capacity, and it should retain a safe fallback ladder such as Normal $\rightarrow$ Low-bitrate $\rightarrow$ AI reconstruction $\rightarrow$ audio-only.  Additionally, the system should be robust to brief failures of the inference service by storing recent reference data, pre-warming models where feasible, and employing timeouts so that errors do not stop the request.

\section{Conclusion}

This study illustrates an adaptive video conferencing solution that stays functional under severe bandwidth restrictions by transitioning from pixel-based video transmission to an audio-driven talking-head reconstruction mode guided by real-time telemetry.  The long-run bandwidth logs show a dramatic reduction in sustained network demand when AI mode is enabled, as summarized in Table~\ref{tab:bandwidth_summary}, indicating that the system can operate in a regime that is far below the typical requirements of conventional video conferencing while still providing a visual stream.  The results framework further defines how performance should be judged under specific bandwidth limitations since the median usage of bandwidth is 32.80kbps, employing stability. From an engineering standpoint, the key contribution is the semantic substitution of video: instead of attempting to compress pixels beyond practical limits, the system transmits speech audio and compact visual control information and reconstructs a plausible talking-head output, which reduces bandwidth demand and improves stability in constrained networks.  The research further highlights that this capacity must be accompanied with rigorous mode-switching logic, secure fallbacks, and responsible privacy and disclosure procedures since synthesized media transforms the security and trust model of communication.  Future work can strengthen the system by improving robustness to pose and occlusion, extending support for multi-speaker and multi-party identity management, reducing synthesis latency on low-power devices, and integrating stronger safeguards and transparency mechanisms suitable for sensitive application domains.

\bibliographystyle{plain}  

\section*{Author contributions}

\textbf{Vineet Kumar Rakesh:} Conceptualization, Methodology, Software (backend development and service deployment), Visualization, Writing (review \& editing), Supervision. \textbf{Soumya Mazumdar:} Software (frontend development and partial contribution to \texttt{JavaScript}), Validation, Investigation, Writing (original draft). \textbf{Tapas Samanta:}  Validation. \textbf{Hemendra Kumar Pandey:} Resources, Investigation, Validation. \textbf{Amitabha Das:} Validation. \textbf{Sarbajit Pal:} Resources, Validation. All authors reviewed and approved the final version of the manuscript and agreed to be accountable for all aspects of the work.

\section*{Acknowledgments}

This research was supported by the Variable Energy Cyclotron Centre (VECC) and the Homi Bhabha National Institute (HBNI), both under the Department of Atomic Energy (DAE), Government of India, which provided essential facilities and technical support. The authors thank the peer reviewers for their insightful comments and constructive suggestions. We also extend our appreciation to the staff of the VECC Library for their valuable assistance throughout the course of this study.

\section*{Declaration of Competing Interest}

The authors state that they own no recognized conflicting financial interests or personal ties that may have seemingly influenced the work presented in this study.

\section*{Data Availability Statement}

The data behind the conclusions of this investigation may be obtained from the corresponding author upon a reasonable request. All findings provided in this publication were produced immediately during the investigation; hence, no extra permissions were necessary to access or use the data.

\section*{Author Biographies}

\AuthorBio{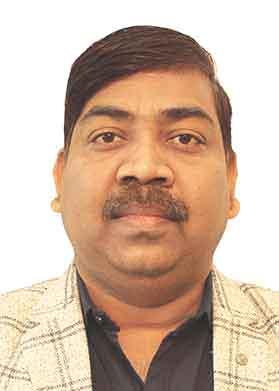}{Vineet Kumar Rakesh}{
is a Technical Officer (Scientific Category) at the Variable Energy Cyclotron Centre (VECC), Department of Atomic Energy, India, with over 22 years of experience in software engineering, database systems, and artificial intelligence. His research focuses on talking head generation, lipreading, and ultra-low-bitrate video compression for real-time teleconferencing. He is pursuing a Ph.D. at Homi Bhabha National Institute, Mumbai. Mr. Rakesh has contributed to office automation, OCR systems, and digital transformation projects at VECC. He is an Associate Member of the Institution of Engineers (India) and a recipient of the DAE Group Achievement Award.
}

\AuthorBio{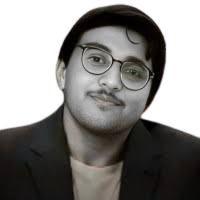}{Soumya Mazumdar}{
is pursuing a dual degree: a B.Tech in Computer Science and Business Systems from Gargi Memorial Institute of Technology, and a B.S. in Data Science from the Indian Institute of Technology Madras. He has contributed to interdisciplinary research with over 25 publications in journals and edited volumes by Elsevier, Springer, IEEE, Wiley, and CRC Press. His research interests include artificial intelligence, machine learning, 6G communications, healthcare technologies, and industrial automation.
}

\AuthorBio{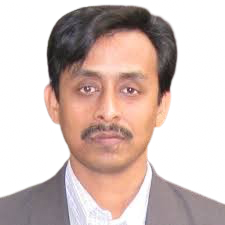}{Dr. Tapas Samanta}{
is a senior scientist and Head of the Computer and Informatics Group at the Variable Energy Cyclotron Centre (VECC), Department of Atomic Energy, India. With over two decades of experience, his work spans artificial intelligence, industrial automation, embedded systems, high-performance computing, and accelerator control systems. He also leads technology transfer initiatives and public scientific outreach at VECC.
}

\AuthorBio{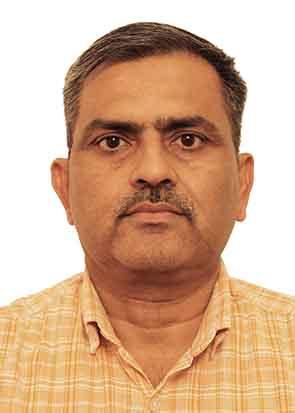}{Hemendra Kumar Pandey}{is a Scientific Officer in the Radioactive Ion Beam Facilities Group at the Variable Energy Cyclotron Centre (VECC), Department of Atomic Energy, Kolkata, India. He received his Ph.D. from the Indian Institute of Technology Kharagpur and his M.Tech. from the University of Allahabad. He joined Bhabha Atomic Research Centre in 1999 and has been associated with VECC since 2000, where he has contributed to RF and microwave systems for particle accelerators, including development activities for the Radioactive Ion Beam facility. He is also an Associate Professor at Homi Bhabha National Institute. His research interests include RF systems for particle accelerators, beam diagnostics, high-power RF amplifier development, mixed-signal RF integrated-circuit design, and radiation-hardened devices.in accelerator-based technologies.
}

\AuthorBio{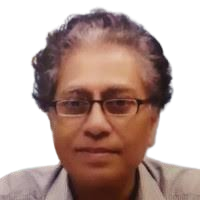}{Dr. Amitabha Das}{
is the Director and Head of the School of Nuclear Studies and Application at Jadavpur University, Kolkata. His research interests include nuclear instrumentation, embedded systems, reactor control systems, and FPGA-based real-time data acquisition. He has also contributed to AI-driven applications such as lipreading and sign language recognition and has supervised advanced research in nuclear reactor control methodologies.
}

\AuthorBio{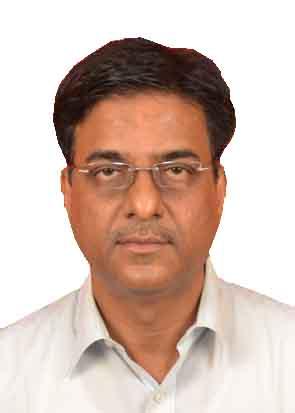}{Dr. Sarbajit Pal}{
is a retired senior scientist and former Head of the C\&I Group at the Variable Energy Cyclotron Centre (VECC), Department of Atomic Energy, Government of India. He holds a Ph.D. in Electronics Engineering and has made significant contributions to control and instrumentation systems for particle accelerators, including the K500 Superconducting Cyclotron. His expertise includes embedded systems, experimental physics, and EPICS-based control architectures.
}

\end{document}